\documentclass[aps,reprint,prl,superscriptaddress,amsmath,amssymb,floatfix]{revtex4-2}

\usepackage[utf8]{inputenc}
\usepackage{xcolor}
\usepackage{graphicx}
\usepackage[colorlinks,allcolors=blue]{hyperref}

\newcommand{\cre}[1]{{b}^\dagger_{#1}}
\newcommand{\ann}[1]{{b}_{#1}}

\newcommand{\num}[1]{{n}_{#1}}

\begin{document}

\title{Bose-Hubbard model with power-law hopping in one dimension}


\author{Tanul Gupta}
\affiliation{University of Strasbourg and CNRS, CESQ and ISIS (UMR 7006), aQCess, 67000 Strasbourg, France}

\author{Nikolay V. Prokof'ev}
\affiliation{Department of Physics, University of Massachusetts, Amherst, MA 01003, USA}

\author{Guido~Pupillo}
\affiliation{University of Strasbourg and CNRS, CESQ and ISIS (UMR 7006), aQCess, 67000 Strasbourg, France}

\begin{abstract}
We investigate the zero-temperature phase diagram of the one-dimensional Bose-Hubbard model with power-law hopping decaying with distance as $1/r^\alpha$ using exact large scale quantum Monte Carlo simulations. For all $1<\alpha\leq 3$ the quantum phase transition from a superfluid and a Mott insulator at unit filling is found to be continuous and scale invariant, in marked contrast with the Berezinskii–Kosterlitz–Thouless (BKT) scenario that is recovered only for $\alpha>3$. By performing finite-size scaling collapses of the superfluid stiffness and extracting dynamical and correlation-length exponents from the low-energy spectrum, we establish that these transitions define a distinct universality class throughout the long-range regime $1<\alpha\le 3$. Analysis of the single-particle correlation functions and grand canonical phase diagram further reveals a sequence of ordering regimes within the superfluid phase: true long-range order for $\alpha\le 2$, anomalous quasi–long-range order for $2<\alpha\le 3$, and conventional algebraic decay for $\alpha>3$. Our exact numerical results provide a benchmark to compare theories of long-range quantum models and are relevant for experiments with cold neutral atom, molecules and ion chains. 
\end{abstract}

\maketitle
The Bose–Hubbard (BH) model describes the dynamics of interacting bosons on a lattice with nearest-neighbor hopping $t$ and on-site repulsion $U$. It features a quantum phase transition from a gapless superfluid (SF) to a gapped Mott insulator (MI) as a function of the ratio $t/U$ \cite{Fisher1989, Batrouni19901DBHModel, freericks1994phase, BlochRMP}, and captures localization physics in systems including ultracold atoms in optical lattices, superfluid $^4\mathrm{He}$, and superconductors \cite{Jaksch1998, GreinerExp2002, BlochRMP, Yu2021}. In one dimension with short-range hopping, the SF–MI transition at integer filling belongs to the Berezinskii–Kosterlitz–Thouless (BKT) universality class \cite{giamarchi1987localization, berezinskii1972destruction, Kosterlitz1973, Kosterlitz1974, BKT40years, GiamarchiBook}, a hallmark of one-dimensional quantum criticality confirmed in cold-atom experiments \cite{Nagerl2010, Boeris2016}. This BKT scenario is widely regarded as the generic mechanism underlying all localization transitions in one-dimensional bosonic systems. 

\begin{figure}[t]
\includegraphics[width=\columnwidth]{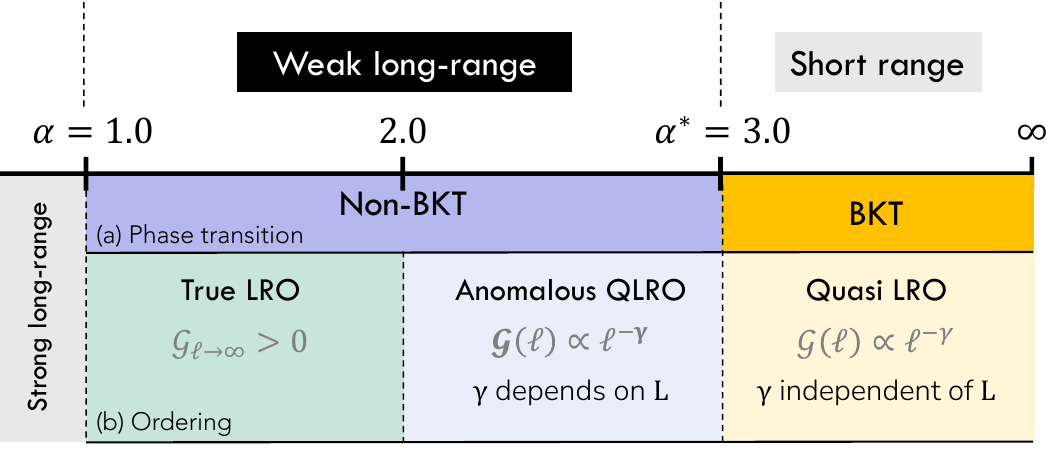}
    \caption{Summary of quantum critical behavior and ordering regimes in the 1D Bose–Hubbard model with power-law hopping $t_{ij}\propto 1/r_{ij}^{\alpha}$. (a) For $1<\alpha\le3$, the SF-MI transition is continuous, scale-invariant, and non-BKT. For $\alpha > 3$, it belongs to the BKT universality class, identifying $\alpha_* = 3$. (b) Within the SF phase, decay of the single-particle density matrix demonstrate three regimes: (i) true long-range order (LRO) for $\alpha\le2$, (ii) anomalous quasi-long-range order (AQLRO) for $2<\alpha\le3$, where the decay exponent $\gamma$ depends on $\alpha$ and the system size $L$, and (iii) conventional quasi-long-range order for $\alpha>3$, where correlations decay algebraically with an $L$-independent exponent (see Fig. \ref{fig:4OBDM}).}
\label{fig:1PPD}
\end{figure}

An important open question is how this picture is altered when the hopping acquires a long-range character. This has become experimentally relevant with recent advances in quantum simulators — including Rydberg atom arrays, dipolar atoms and molecules, trapped-ion chains, and atoms coupled to photonic or cavity modes \cite{browaeysExperimentalInvestigationsDipole2016, baranovCondensedMatterTheory2012, lahayePhysicsDipolarBosonic2009, britton2012engineered, Islam2013} — which now enable controlled realization of long-range hopping with tunable power-law decay. These platforms have revealed many-body phenomena beyond short-range paradigms \cite{DefenuLongRange2023}, such as non-local propagation of correlations \cite{haukeSpreadCorrelationsLongRange2013, schachenmayerEntanglementGrowthQuench2013, eisertBreakdownQuasilocalityLongRange2013, Richerme2014, Jurcevic2014}, breakdown of conformal symmetry \cite{vodolaKitaevChainsLongRange2014, leporiEffectiveTheoryBreakdown2016}, and long-range-induced novel topological phases and phase transitions \cite{vodolaKitaevChainsLongRange2014, gongLongRangeFreeFermions2023, leporiLongrangeTopologicalInsulators2017}.

These observations raise the question of how long-range hopping, in particular hopping that decays algebraically as $t_{ij} \propto 1/r_{ij}^{\alpha}$, modifies the Bose–Hubbard phase diagram. Bose–Hubbard models with such power-law hopping have recently been realized using indirect dipolar excitons in GaAs bilayers \cite{arxivLagoin2024BHLRH, arxivMorinSupersolidCrystalsDipolar2025}. Long-range hopping, though not strictly algebraic, can also be engineered in superconducting-circuit platforms using transmon qubits embedded in photonic-bandgap metamaterial waveguides \cite{Sundaresan2019, zhangSuperconducting2023}.

The most interesting regime is the weak long-range regime, defined by $d<\alpha<\alpha_*$, where $d$ is the spatial dimension, $\alpha$ controls the decay of hopping or interactions, and $\alpha_*$ denotes the threshold separating long-range and effectively short-range critical behavior. In this regime, the hopping decays slowly enough to reshape the low-energy dispersion and correlation structure, yet not so slowly as to enforce the mean-field–like, effectively infinite-range behavior occuring for $\alpha \le d$ \cite{DefenuLongRange2023}. Despite significant theoretical progress and exact results in certain integrable limits \cite{vodolaKitaevChainsLongRange2014}, the nature of quantum phases and phase transitions in non-integrable long-range models remains poorly understood. It is thus of fundamental importance to obtain exact results for these systems, to which theories can be compared.

In this work, we determine the ground-state phase diagram of the one-dimensional Bose–Hubbard model with power-law hopping $t_{ij} \propto 1/r_{ij}^{\alpha}$ using numerically exact large-scale quantum Monte Carlo simulations based on the Worm Algorithm \cite{prokof1998exact}. Interestingly, at unit filling, finite-size scaling of the winding-number fluctuations—which directly probe superfluid stiffness—shows that the superfluid–Mott-insulator transition is incompatible with the BKT universality class for any $1 < \alpha \le 3$. Instead, the transition is continuous and scale-invariant, with critical exponents obtained from data collapse at the critical point and a sublinear low-energy dispersion extracted from the single-particle Green’s function (see Supplemental Material ~\cite{Note1}). For $\alpha > 3$, the transition crosses over to the conventional BKT form, identifying $\alpha_* = 3$ as the boundary between long- and short-range regimes. Figure \ref{fig:1PPD} summarizes these results. Within the superfluid phase, analysis of the single-particle density matrix reveals three distinct regimes: true long-range order for $\alpha \le 2$, anomalous quasi-long-range order for $2 < \alpha \le 3$, and conventional algebraic quasi-long-range order for $\alpha > 3$. These results demonstrate the existence of a distinct universality class of long-range quantum criticality in one dimension and provide exact benchmarks for analytical approaches and quantum simulation experiments.

The $1d$ BH model with power-law hopping reads 
\begin{equation}\label{eq:hamiltonian}
    \mathcal{H} =
    -t \sum_{i < j} \frac{a^\alpha}{|r_{ij}|^\alpha}
    \left[ \cre{i}\ann{j} + \text{H.c.} \right]
    + \frac{U}{2}\sum_i \num{i}(\num{i}-1) - \mu \sum_i\num{i}
\end{equation}
Here, $b_i^\dagger$, $b_i$, and $n_i = b_i^\dagger b_i$ are the bosonic creation, annihilation, and number operators on site $i$, respectively. The parameters $t$, $U$, and $\mu$ denote the hopping amplitude, on-site interaction strength, and chemical potential, while $a$ is the lattice spacing. Throughout, we set the energy and length scales by choosing $t = 1$ and $a = 1$. For nearest-neighbor hopping $(\alpha \rightarrow \infty)$ at unit filling, Eq. \eqref{eq:hamiltonian} exhibits a zero-temperature quantum phase transition of the BKT type between a superfluid and a Mott insulator at a critical value $(t/U)_c = 0.300 \pm 0.025$ \cite{krauth1991bethe, elesin1994mott, kashurnikov1996mott, kuhner1998phases}. For finite $\alpha$ and in the hard-core limit $U/t\rightarrow \infty$, Eq. \eqref{eq:hamiltonian} maps to a long-range XY model, for which spin-wave and semi-analytical renormalization group analyses predict a continuously varying dynamical exponent $z = (\alpha-1)/2$ for $\alpha<3$ \cite{Roscilde2017} as well as a breaking of the $U(1)$ symmetry leading to true long-range order for some $\alpha_c<3$, whose precise value must be determined numerically. In Ref. \cite{maghrebi_gong_gorshkov_2017}, $\alpha_c$ was estimated to be $\alpha_c \simeq 2.8$ using a density matrix renormalization group approach for system sizes up to $L \simeq 100$. In this work, we investigate Eq. \eqref{eq:hamiltonian} for $\alpha> 1$ and arbitrary $t/U$ using large-scale quantum Monte Carlo simulations on chains of up to $L=1024$ sites and inverse temperature $\beta=L^{z}$, small enough to probe ground state properties. We first analyze the superfluid-Mott insulator transition at fixed density, and then characterize the correlation functions in the liquid phase.\\
\begin{figure}[t]
\includegraphics[width=\columnwidth]{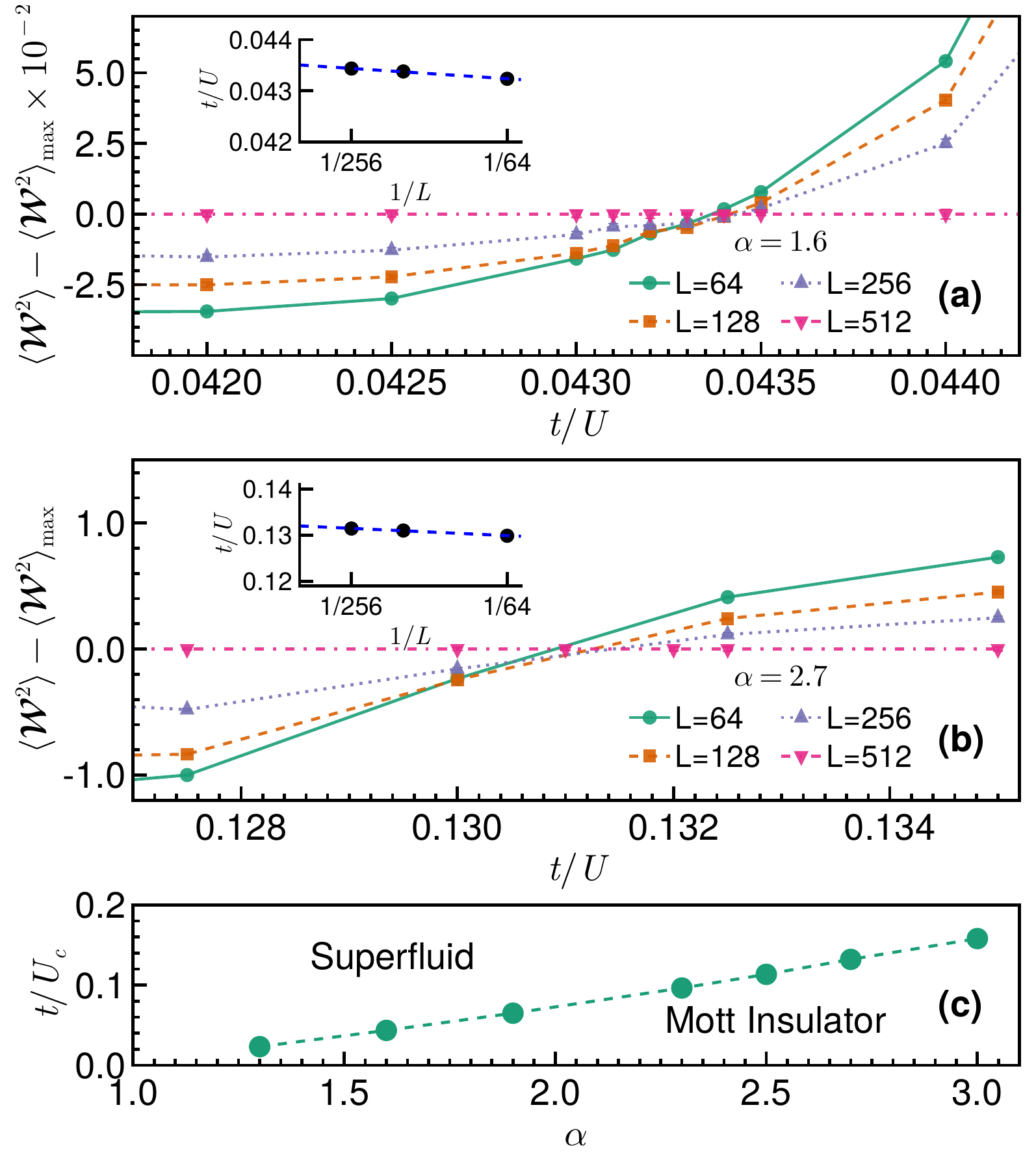}
    \caption{Characterization of the Mott insulator to superfluid phase transition: (a)-(b) Mean-square winding number $\langle \mathcal{W}^2 \rangle$ vs $t/U$ for $\alpha=1.6$ and $2.7$ near phase transitions, showing crossing between the curves; the curve corresponding to the largest lattice size is subtracted from all data for clarity. Vertical error bars indicate the estimated uncertainty from the Monte Carlo simulations. Insets: Finite-size scaling of crossings points between curves for system sizes $L_1$ and $L_2=2L_1$ as a function of $L=L_1$. (c) Phase diagram, ${t/U}_c$ vs $\alpha$ of the Mott insulator and superfluid quantum phases for model \eqref{eq:hamiltonian}.}
    \label{fig:2Winding}
\end{figure}

\emph{Quantum phase transition.—} To characterize the SF-MI transition at unit filling $\rho = 1$, we compute the mean-squared winding number fluctuations $\langle \mathcal{W}^2 \rangle$—a scale-invariant quantity proportional to the superfluid stiffness $Y_s$ as $\langle\mathcal{W}^2\rangle = Y_s/(LT)$— for system sizes up to  $L=512$. In the superfluid phase, $\langle \mathcal{W}^2 \rangle$ remains finite, while it vanishes in the Mott insulating phase, providing a sharp diagnostic of the transition. Figures~\ref{fig:2Winding}(a) and (b) show $\langle \mathcal{W}^2 \rangle$ as a function of $t/U$ for two power-law exponents $\alpha < 2$ and $\alpha > 2$ in Eq. \eqref{eq:hamiltonian}, respectively, and for several system sizes $L$. For clarity, the values of $\langle \mathcal{W}^2 \rangle_{L_\text{max}}$at the largest system size $L_{\mathrm{max}}$ used in the computations have been subtracted. In both cases, the curves for different $L$ display a clear crossing at $(t/U)_c = 0.043 \pm 0.001$ and $(t/U)_c = 0.131 \pm 0.001$, respectively (see insets for extrapolation to the thermodynamic limit). These crossings correspond to a quantum phase transition at the respective values of the critical ratio $(t/U)_c$. The very presence of such crossings directly rules out BKT universality class for these power-law models, in contrast to both short-range one-dimensional systems (see Supplemental Material ~\cite{Note1}) and long-range models with power-law density-density interactions \cite{botzungEffectsEnergyExtensivity2021, DalmontePupilloPRL}. We observe similar crossings for all $1 < \alpha \le 3$, demonstrating a continuous, scale-invariant quantum phase transition throughout this range. For $\alpha > 3$, the transition reverts to the conventional BKT form (see Supplemental Material), indicating that $\alpha_\ast = 3$ marks the onset of  short-range critical behavior.

Figure \ref{fig:2Winding}(c) shows the ground state phase boundary of Eq. \eqref{eq:hamiltonian} between the SF and MI ground states as a function of $(t/U)_c$ and $1<\alpha\leq 3$. In this diagram, $(t/U)_c$ for each value of $\alpha$ is identified based on the scale-invariant crossing point, as previously detailed. The discovery of this family of scale-invariant phase transitions is a key result of this work.

\begin{figure}[t]
    \centering
    \includegraphics[width=\columnwidth]{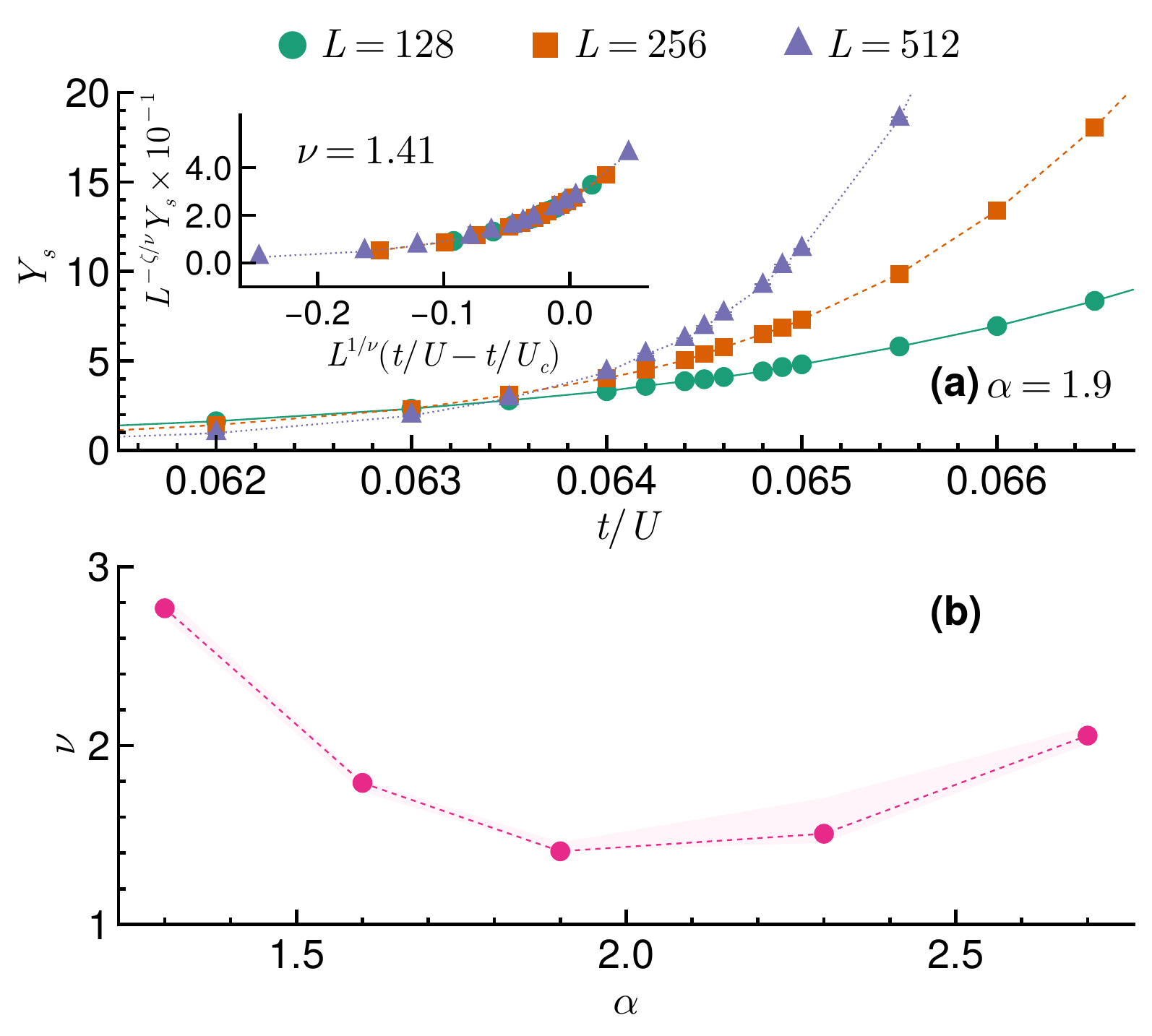}
    \caption{(a) Superfluid stiffness $Y_s$ near the critical interaction strength ${t/U}_c$ for $\alpha=1.9$ and system sizes $L=128, 256, 512$. Inset: Finite-size scaling collapse for the rescaled stiffness $L^{-\zeta/\nu}Y_s$ as a function of $L^{1/\nu}[(t /U) - ({t/U}_c]$, from which the correlation length exponent $\nu$ is extracted.
(b) Correlation-length exponent $\nu$ obtained from scaling collapses as a function of the hopping exponent $\alpha$. The minimum near $\alpha \simeq 2$ indicates a qualitative change in the critical behavior.}
    \label{fig:3FSSA}
\end{figure}

We further characterize the SF-MI quantum phase transitions at commensurate density by determining the correlation length exponent $\nu$ associated with the continuous transition using data collapse analysis near the critical points. For each $\alpha$, we rescale the superfluid stiffness using  $L^{-\zeta/\nu} Y_s$  as a function of  $L^{1/\nu}[(t /U) - (t/U)_c]$ , where  $\nu$  and  $\zeta$  are fitting parameters. These parameters are determined through optimization using the Nelder-Mead algorithm \cite{Nelder1965}, with a cost function based on the Kawashima-Ito-Houdayer-Hartmann quality metric \cite{Kawashima1993, Houdayer2004}. An example result for $\alpha = 1.9$ is shown in Fig.~\ref{fig:3FSSA}(a), demonstrating excellent data collapse near the critical point, consistent with scale-invariant critical behavior. The resulting correlation-length exponents $\nu$ for all $1<\alpha<3$ are presented in Fig.~\ref{fig:3FSSA}(b) as a function of the hopping exponent $\alpha$. Notably, $\nu(\alpha)$ exhibits a pronounced minimum near $\alpha\simeq 2$, signalling a qualitative change in the critical behavior and motivating a closer examination of correlation functions in the superfluid phase.\\

\begin{figure*}[t]
\includegraphics[width=\linewidth]{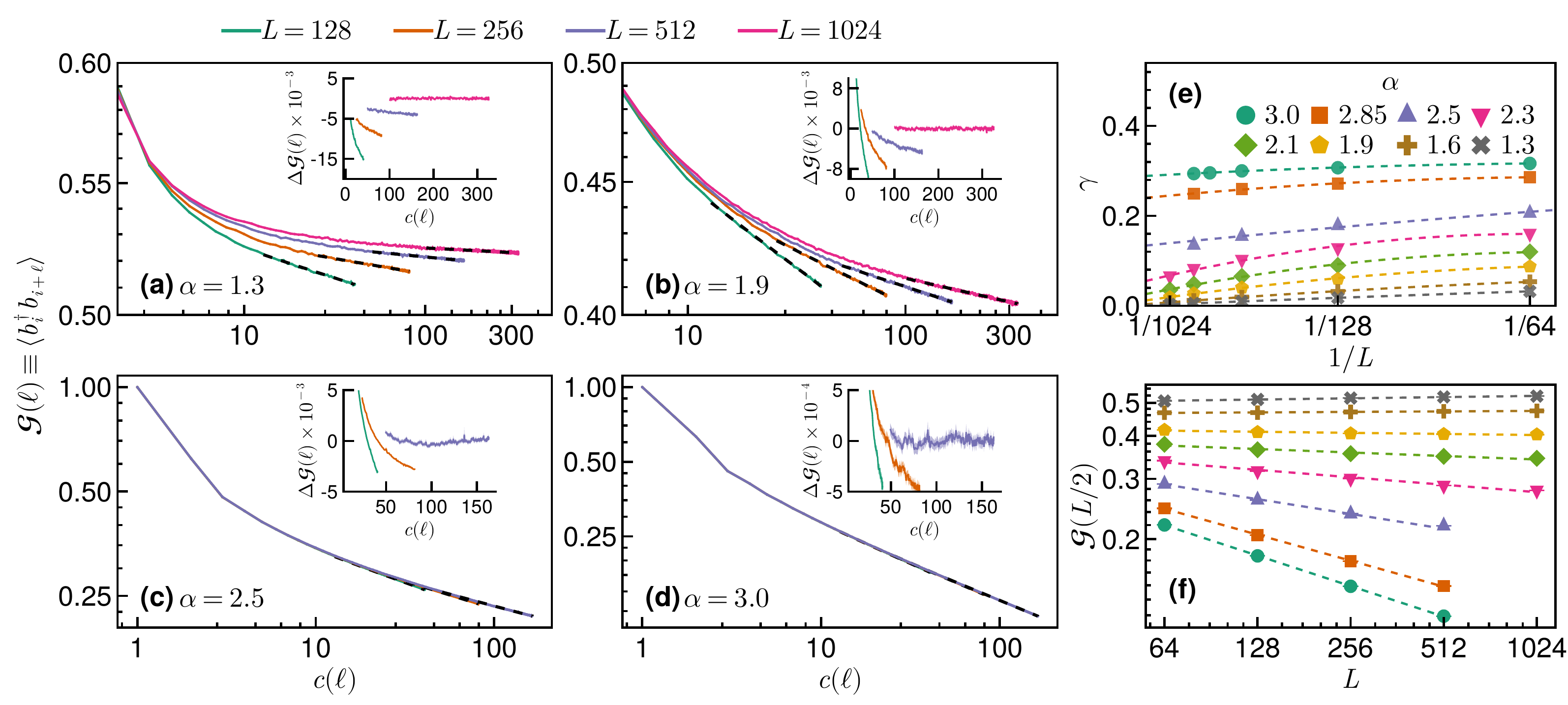}
    \caption{Characterization of the superfluid phase: (a)-(d) Single-particle density matrix, $\mathcal{G}(\ell)$, plotted against $\ell$ for $\alpha = 1.3$, $1.6$, $2.5$, and $2.7$. The dashed line represents the best fit to $A \cdot c(\ell)^{-\gamma}$, where $c(\ell)=\sin(\pi \ell/L)/\sin(\pi/L)$ is the chord distance, and $A$ and $\gamma$ are fitting parameters. For $\alpha = 1.3$ and $1.9$, $\mathcal{G}(\ell)$ saturates to a constant as $\ell \rightarrow \infty$. For $\alpha = 2.5$ and $2.7$, $\mathcal{G}(\ell)$ exhibits algebraic decay. Insets: Difference $\Delta \mathcal{G}(\ell)$ between $\mathcal{G}(\ell)$ for system size $L$ and the numerical fit for $L = L_{\text{max}}$, plotted as a function of the chord distance $c(\ell)$. The difference indicates an upward trend of $\mathcal{G}(\ell)$ for increasing lattice size. (e) Finite-size scaling of the power-law exponent $\gamma$ to the thermodynamic limit. (f) Condensate fraction $\mathcal{G}(L/2)$ as a function of system size $L$.
    }
    \label{fig:4OBDM}
\end{figure*}

\emph{Single-particle correlations and ordering.—} To probe the nature of correlations inside the superfluid phase, we examine the single-particle density matrix $\mathcal{G}(\ell) = \langle\, b_i^\dagger b_{i+\ell}\,\rangle$ as a function of distance $\ell$ near the critical point. Example results are shown in Fig.~\ref{fig:4OBDM}(a–d), plotted against the chord distance $c(\ell)=\sin(\pi \ell/L)/\sin(\pi/L)$, which accounts for periodic boundary conditions. The data correspond to the hard-core limit $(U/t\to\infty)$ at half filling, which allows simulations up to $L=1024$ and facilitates direct comparison with analytical predictions for the long-range XY model \cite{maghrebi_gong_gorshkov_2017, dupuis2024}. For all $1<\alpha\le 3$, $\mathcal{G}(\ell)$ decays monotonically with distance, and for each $\alpha$ and finite $L$, the decay is extremely well described by a power law $\mathcal{G}(\ell)\sim c(\ell)^{-\gamma}$ (see Supplemental Material). 

In sharp contrast to the short-range one-dimensional systems—where true long-range order is forbidden and correlations decay algebraically with an $L$-independent exponent as described by Luttinger-liquid theory and the Hohenberg–Mermin–Wagner theorem \cite{GiamarchiBook, MerminWagner, hohenberg1967}—the effective decay exponent $\gamma$ here exhibits a pronounced system size dependence. Specifically, $\gamma$ decreases as $L$ increases, indicating nontrivial finite-size scaling and raising the question of its asymptotic value in the thermodynamic limit. The fitted decay exponents $\gamma(L)$ are plotted as a function of $1/L$ in Fig.~\ref{fig:4OBDM}(e). Extrapolation to the thermodynamic limit shows that $\gamma \to 0$ for $1<\alpha\lesssim 2$, whereas $\gamma$ remains finite for $2\lesssim \alpha \le 3$. The vanishing of $\gamma$ for $1<\alpha\lesssim 2$ indicates the presence of true long-range order (LRO), consistent with theoretical expectations for related long-range interacting models \cite{DefenuLongRange2023, maghrebi_gong_gorshkov_2017, laflorencieCriticalPhenomenaQuantum2005a}. For $2\lesssim \alpha \le 3$, the data can alternatively be fitted using a combination of a power-law decay and an $L$-dependent constant (see Supplemental Material), but the pure power-law form provides a more stable fit with fewer parameters. Because the resulting $\gamma$ values are small but finite in this regime, we refer to $2\lesssim\alpha\le 3$ as exhibiting anomalous quasi-long-range order (AQLRO), distinct from the conventional quasi–long-range order (QLRO) of short-range models. 


This conclusion is further supported by examining the finite-size scaling of the density matrix at the largest separation, $\mathcal{G}(L/2)$, shown in Fig.~\ref{fig:4OBDM}(f). Since $\mathcal{G}(L/2)$ directly measures the condensate fraction in finite systems, it is expected to approach an $L$-independent constant in the presence of long-range order and to decay algebraically with $L$ in a quasi–long-range-ordered phase. Figure \ref{fig:4OBDM}(f) shows that $\mathcal{G}(L/2)$ saturates to a constant for $\alpha<2$ and decays as a power-law with increasing $L$ for $\alpha >2$. We find no evidence to support a speculation that this behavior will change at length scales larger than what were computed here. In the Supplemental Material, we provide additional data showing that soft-core bosons up to $L=512$ exhibit the same scaling behavior of $\mathcal{G}(\ell)$ in the superfluid phase. In the Mott insulating phase, $\mathcal{G}(\ell)$ instead decays as a power law, with the decay exponent $\gamma$ converging to the hopping exponent $\alpha$ in the thermodynamic limit, consistent with the expected behavior of localized bosons with power-law hopping \cite{tgupta2023}.

\begin{figure}[t]
\includegraphics[width=\columnwidth]{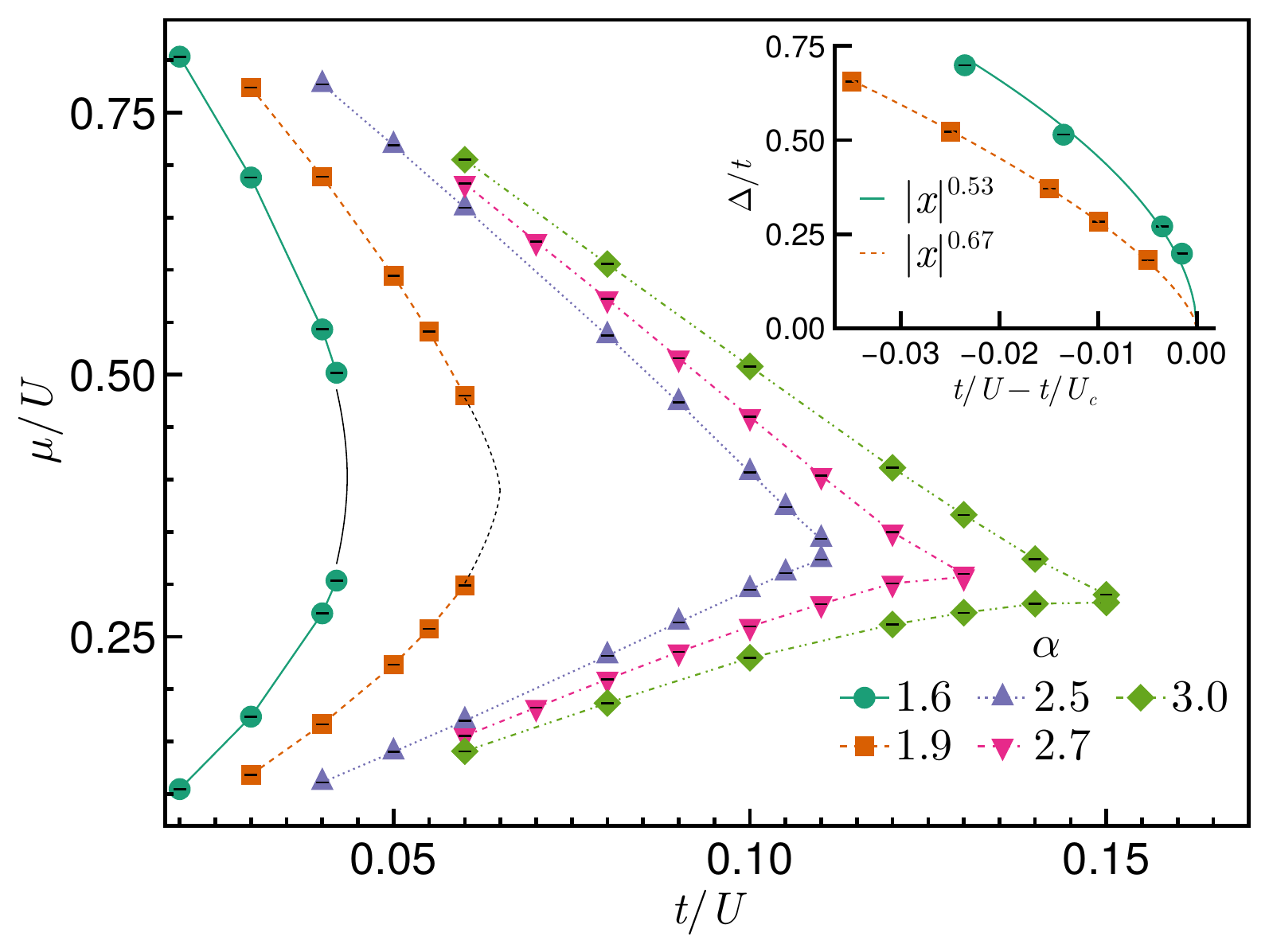}
    \caption{Ground state phase diagram of the 1D Bose-Hubbard model with power-law hopping for $\alpha = 1.6$, $1.9$, $2.5$, $2.7$, and $3.0$, showing the boundary between the Mott insulator (at unity filling) and superfluid phases as a function of chemical potential $\mu/U$ and hopping amplitude $t/U$. Markers represent simulation data, while the solid black line (for $\alpha = 1.6$) and the dashed black line (for $\alpha = 1.9$) are obtained from fitting the energy gap $\Delta/t$ near the critical point as $\Delta/t \sim |t/U - {t/U}_{\rm c}|^{-z\nu}$. Inset: Gap Energy $\Delta/t$ fitted as a function of $x\equiv t/U - {t/U}_{\rm c}$.}
\label{fig:4PhaseDiagram}
\end{figure}

Finally, we examine how the change in ground-state properties at $\alpha \simeq 2$ manifests itself in the  phase diagram. Figure~\ref{fig:4PhaseDiagram} shows the phase diagram in the $(\mu/U,\, t/U)$ plane for several values of $\alpha \le 3$. For each $\alpha$, the diagram displays a Mott insulating (MI) lobe at unit filling, surrounded by a superfluid (SF) phase. The boundaries of the MI region are obtained by extracting the particle and hole excitation energies from the long-time behavior of the zero-momentum Green function $G(k=0, \tau)$, which is obtained via the spatial averaging of the Matsubara Green function $G(i, \tau) = \langle b_i^\dagger(\tau) b_0(0) \rangle$, with  $\tau \in [-\beta/2, \beta/2]$ the imaginary time and $k$ the quasi momentum. Using the Lehmann representation, $G(k,\tau)$ decays exponentially for $|\tau|\to\infty$ as $G(k,\tau)\propto e^{-\epsilon_\pm(k)|\tau|}$, where $\epsilon_{+}$ and $\epsilon_{-}$ are the particle and hole excitation energies, respectively \cite{boninsegni2006fate, capogrosso2007phase}. Fitting the asymptotic tails for $\tau>0$ and $\tau<0$ yields $\epsilon_{+}$ and $\epsilon_{-}$, from which the insulating gap follows as $\Delta = \mu_{+}-\mu_{-}$, with $\mu_{\pm} = \mu \pm \epsilon_{\pm}$ defined relative to the simulation chemical potential $\mu$.

The resulting phase boundaries exhibit a shrinking of the MI lobes in the $t/U-\mu/U$, consistent with earlier predictions for power-law hopping  \cite{ferraretto2019}. Beyond this overall compression, the lobe geometry undergoes a clear qualitative change across $\alpha\simeq2$. For $\alpha < 2$, the MI lobes are smooth and rounded, closely resembling those of higher-dimensional Bose–Hubbard systems \cite{capogrosso2008monte, capogrosso2007phase}; this resemblance, however, reflects a suppression of quantum fluctuations induced by long-range hopping rather than any effective increase in dimensionality. In contrast, for $\alpha > 2$, the lobes become increasingly pointed and asymmetric, and the critical point at unit filling (i.e. the tip of the former lobe) shifts to larger $t/U$, approaching the needle-shaped structure known from the short-range one-dimensional model exhibiting a BKT transition (which is recovered for $\alpha>3$). Together with the qualitative change in the decay of the single particle density matrix $\mathcal{G}(\ell)$ and the pronounced minimum in $\nu(\alpha)$ near the transition point, the change in the lobe geometry indicates $\alpha \simeq 2$ as the boundary separating a regime with genuine long-range coherence $(\alpha \le 2)$ from one with anomalous quasi–long-range order $(2 < \alpha \le 3)$. 

As a consistency check of our analysis of the model Eq. \eqref{eq:hamiltonian}, we note that we can estimate the correlation-length exponent $\nu$ also from the closing of the insulating gap in the MI phase using the expression $\Delta/t \sim |t/U - (t/U)_c|^{-z\nu}$, where the dynamical exponent $z$ is extracted from the low-energy dispersion (see supplemental material). We obtain results for $\nu$ fully consistent with those obtained from the finite-size scaling collapse near the SF-MI transition (see discussion above).
For example, we find $\nu = 1.77 \pm 0.03$ at $\alpha=1.6$ and $\nu = 1.49 \pm 0.04$ at $\alpha=1.9$, consistent with values of Fig.~\ref{fig:3FSSA}. 

In summary, our results depart from existing approximate treatments based on bosonization and medium-scale numerical studies \cite{maghrebi_gong_gorshkov_2017} by establishing that long-range hopping modifies the critical behavior throughout the weak long-range regime $1 < \alpha \le \alpha_* = 3$. This is qualitatively distinct from one-dimensional models with short-range hopping and long-range density-density interactions $n_i n_j/r_{ij}^{\alpha}$ \cite{DalmontePupilloPRL, Schulz1993, botzungEffectsEnergyExtensivity2021}, where the long-range tail does not alter the underlying critical description of the gapless phase, which remains Luttinger-liquid–like for $\alpha>1$, while the Coulomb limit $\alpha=1$ exhibits Wigner-crystal-like charge correlations  \cite{Schulz1993}. Here, by contrast, throughout $1 < \alpha < 3$, we uncover and characterize a distinct universality class of the SF–MI transition that is incompatible with the standard BKT scenario. Within this regime, correlations exhibit true long-range order for $\alpha < 2$, while for $2 < \alpha \le 3$ the system realizes an anomalous quasi–long-range-ordered phase, characterized by a system-size-dependent decay exponent and a vanishing condensate fraction only in the thermodynamic limit. For $\alpha > 3$, the system reverts to conventional short-range behavior, exhibiting quasi-long-range order with an $L$-independent decay exponent $\gamma$, consistent with the standard BKT scenario and with recent results on disorder-driven localization in one dimension \cite{tgupta2023}.


Our predictions for the superfluid dispersion and correlation functions are directly testable in current experimental platforms, including dipolar atoms and molecules ($\alpha=3$) \cite{lahayePhysicsDipolarBosonic2009, GreinerExp2002, FerlainoEBHM2016, KetterleBEC2002, chomazDipolarPhysicsReview2023}, trapped-ion quantum simulators $(1<\alpha\lesssim 3)$ \cite{britton2012engineered, Richerme2014, Jurcevic2014}, and subradiant dipolar excitons in semiconductor quantum wells \cite{arxivLagoin2024BHLRH,lagoinMottInsulatorStrongly2022}. Our work provides exact results to benchmark theories for long-range quantum models and opens up multiple other research directions, including the nature of the groundstate in higher dimensions.

{\it Acknowledgments:} T.G. and G.P. gratefully acknowledge discussions with Guido Giachetti. 
This research has received funding from the
European Union’s Horizon 2020 research and innovation programme under the Marie Sklodowska-Curie
project 955479 (MOQS), the Horizon Europe programme
HORIZON-CL4-2021-DIGITAL-EMERGING-01-30 via
the project 101070144 (EuRyQa) and from the French
National Research Agency under the Investments of the
Future Program projects ANR-21-ESRE-0032 (aQCess),
ANR-22-CE47-0013-02 (CLIMAQS) and ANR-23-CE30-
50022-02 (SIX). NP acknowledges support from the National Science Foundation under Grant No. DMR2335904.


\footnotetext[1]{See Supplemental Material for the extraction of the single-particle dispersion relation in the superfluid phase, the analysis of the one-body density matrix and its asymptotic scaling, results for soft-core bosons in both the superfluid and Mott-insulating phases, short-range benchmarks using winding-number fluctuations and Weber-Minnhagen scaling, and a summary of experimental platforms realizing long-range hopping. The Supplemental Material includes Refs.~\cite{Roscilde2017, weberMonteCarloDetermination1988, arxivLagoin2024BHLRH, Richerme2014, Jurcevic2014, Monroe2021, LewisIonTrapLongrange2023, Barredo2015, browaeys_lahaye_2020, emperaugerTomonagaLuttingerLiquidBehavior2025, chen2023continuous, Yan2013, hazzard2014, christakis2023, Sundaresan2019, zhangSuperconducting2023, steinertSpatiallyTunableSpin2023}.}

\bibliography{references}

\end{document}


\title{\vspace{10cm}Supplemental Material for ``Bose-Hubbard model with power-law hopping in one dimension''}

\author{Tanul Gupta}
\affiliation{University of Strasbourg and CNRS, CESQ and ISIS (UMR 7006), aQCess, 67000 Strasbourg, France}

\author{Nikolay V. Prokof'ev}
\affiliation{Department of Physics, University of Massachusetts, Amherst, MA 01003, USA}

\author{Guido~Pupillo}
\affiliation{University of Strasbourg and CNRS, CESQ and ISIS (UMR 7006), aQCess, 67000 Strasbourg, France}

\maketitle

This Supplemental Material provides the technical details and auxiliary analyses supporting the main results on the Bose–Hubbard model with power-law hopping in one dimension. Section \ref{secS1:Dispersion} presents the numerical extraction of the single-particle excitation spectrum in the superfluid regime and its low-momentum dispersion $E(k)$ for $1<\alpha<3$.  In Sec. \ref{secS2:FittingForm}, we analyze the asymptotic behavior of the one-body density matrix and compare different fitting forms used to characterize long-distance correlations. In Sec. \ref{secS3:Softcore} we extend the discussion to soft-core bosons and document the behavior of Green’s functions in both the superfluid (SF) and Mott-insulating (MI) phases. Section \ref{secS4:BKT} provides short-range benchmark results, including winding-number diagnostics and Weber–Minnhagen scaling, to contrast BKT behavior with the long-range regime studied in the main text. Finally, Sec. \ref{secS5:Experiments} briefly summarizes experimental platforms where effective long-range hopping can be realized.

\section{Dispersion Relation in the Superfluid Phase}
\label{secS1:Dispersion}
\begin{figure}[h]
\includegraphics[width=0.5\columnwidth]{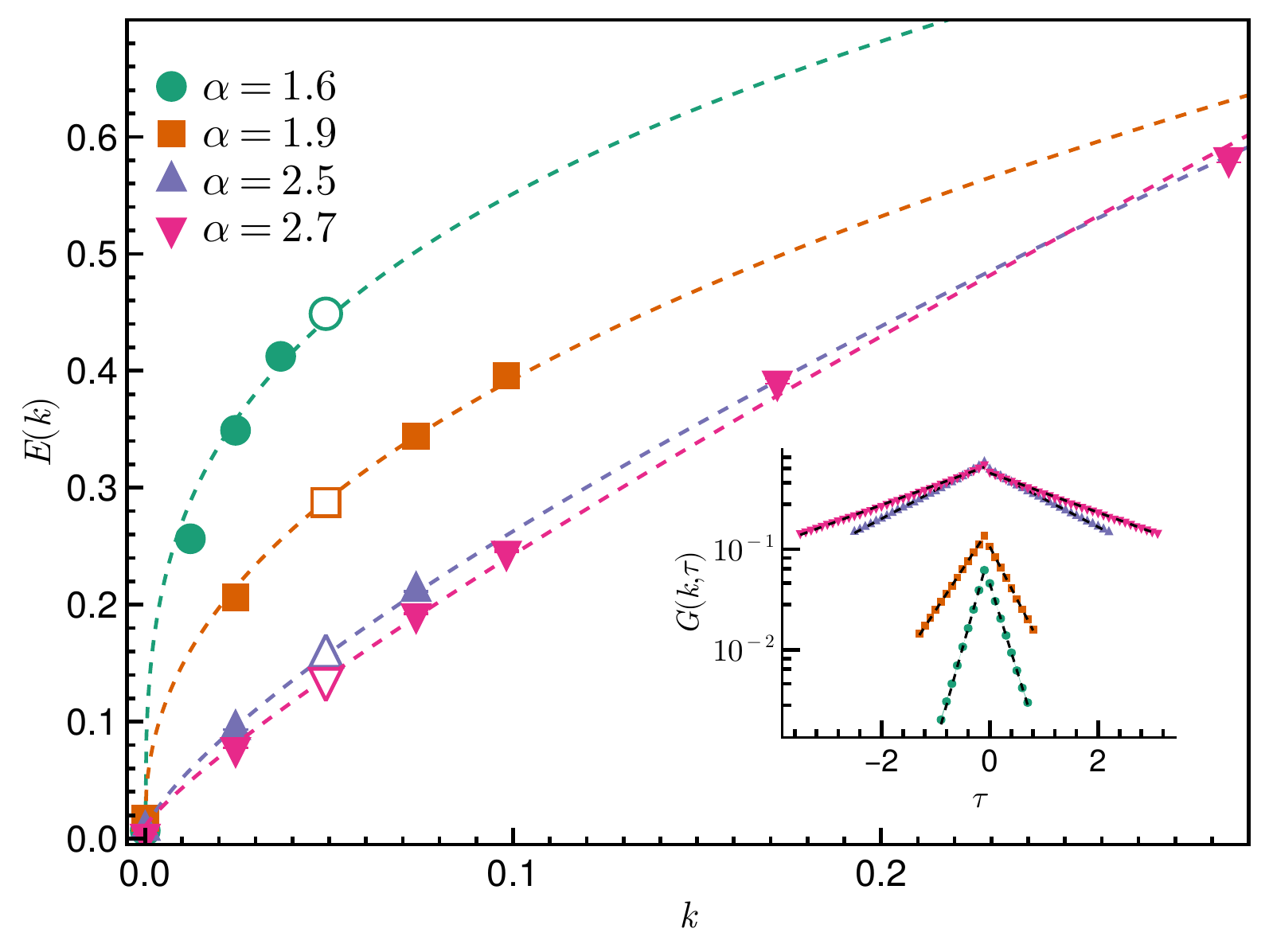}
    \caption{Dispersion relation $E(k)$ vs. $k$ for $\alpha = 1.6$, $1.9$, $2.5$, and $2.7$ in the superfluid phase near the MI-SF transition along the commensurate density line. Symbols denote numerical data obtained from fitting the long-time exponential decay of $G(k, \tau)$, while lines show the spin-wave analysis prediction $E(k) \sim k^{z}$, with $z = (\alpha - 1)/2$ \cite{Roscilde2017}. Inset: Single-particle Green function $G(k, \tau)$ for $k=\pi/64$ (represented by empty markers) and $\alpha=1.6, 1.9, 2.5,$ and $2.7$ for $L=256$. The black dashed line represents the numerical fit to the exponential decay.
    }
    \label{fig:1Spectra}
\end{figure}
The single-particle excitation spectrum in the superfluid phase is characterized by the dispersion relation $E(k)$. Figure \ref{fig:1Spectra} shows $E(k)$ extracted for several values of $1<\alpha<3$ at constant density near the transition point. The quantum Monte Carlo estimates (symbols) are obtained by fitting the long-$\tau$ decay of the momentum-resolved Green’s function $G(k,\tau)$ to an exponential form, $G(k,\tau)\sim e^{-\epsilon\pm(k)|\tau|}$ (see inset for representative fits). Here, $\epsilon_+(k)$ and $\epsilon_-(k)$ denote the particle and hole excitation energies, respectively, and the dispersion is given by $E(k)=\epsilon_+(k)-\epsilon_-(k)$. The resulting dispersions are sublinear in momentum $k$, with increasing curvature as $\alpha$ decreases. The dashed lines indicate power-law fits $E(k)\sim k^{z}$, where $z$ is the dynamical critical exponent describing the data accurately in the low-momentum regime. The extracted exponents are consistent with the spin-wave prediction $z=(\alpha-1)/2$ \cite{Roscilde2017}, indicating that the dynamical exponent varies continuously with $\alpha$ within the window $1<\alpha<3$.

\section{One-body density matrix: power-law vs offset fits}
\label{secS2:FittingForm}

\begin{figure}[h]
\includegraphics[width=\linewidth]{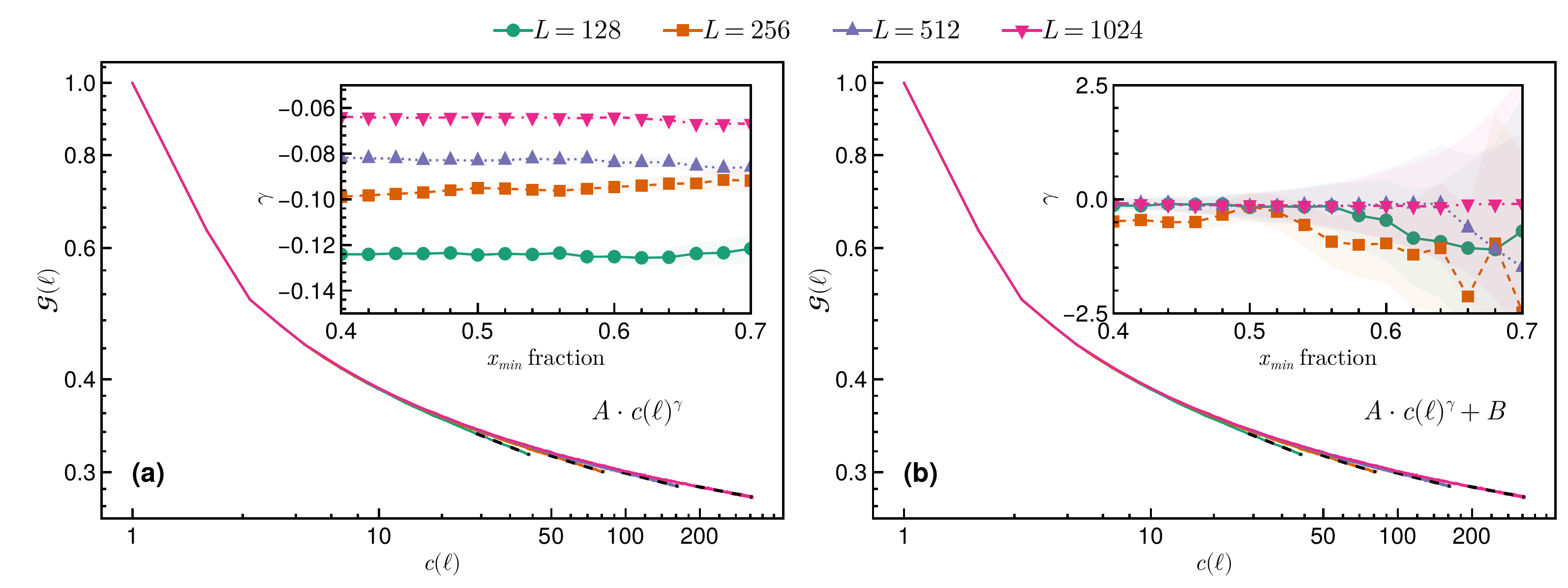}
    \caption{One-body density matrix, $\mathcal{G}(\ell)$, plotted against $c(\ell)$ for $\alpha = 2.3$. (a) The dashed black line represents the power-law fit $A\cdot c(\ell)^{-\gamma}$ to $\mathcal{G}(\ell)$. Inset: Power-law exponent $\gamma$ estimated by fitting $\mathcal{G}(\ell)$ from $\ell=x_{\rm min}$ to $\ell=L/2$ where $x_{\rm min}=x_{\rm min}{\; \rm fraction}\times L/2$. (b) The dashed black line represents the power-law fit with additional constant $A \cdot c(\ell)^{-\gamma} + B$ to $\mathcal{G}(\ell)$. Inset: Power-law exponent $\gamma$ estimated by fitting $\mathcal{G}(\ell)$ from $\ell=x_{\rm min}$ to $\ell=L/2$.
    }
    \label{fig:Glfit}
\end{figure}

Complementary to the momentum-space information contained in the dispersion relation, we examine the real-space decay of single-particle correlations using the one-body density matrix (OBDM), $\mathcal{G}(\ell)$. Figure \ref{fig:Glfit} shows the OBDM $\mathcal{G}(\ell)$ as a function of the chord distance $c(\ell) = \sin(\pi \ell / L)/ \sin(\pi/L)$ for $\alpha = 2.3$, together with a comparison of two fitting forms. In panel (a), the data are fitted to a pure power-law form $A\,c(\ell)^{-\gamma}$, and the inset shows the corresponding exponent $\gamma$ extracted by fitting $\mathcal{G}(\ell)$ over the interval $\ell \in [x_{\rm min}, L/2]$, where $x_{\rm min} = x_{\rm min}^{\rm fraction}\times (L/2)$ defines the lower cutoff as a fraction of the system size. Panel (b) shows the result of fitting the same data to an extended form $A\,c(\ell)^{-\gamma} + B$, where B is a constant offset. The insets display the fitted exponent $\gamma$ as a function of the fitting window for both models. While both models describe the overall behavior of $\mathcal{G}(\ell)$ comparably well, the pure power-law fit is significantly more stable: it yields consistent values of $\gamma$ across a wide range of fitting windows, with $x_{\rm min}$ fractions between 0.2 and 0.8, indicating robust extraction of the asymptotic exponent. In contrast, inclusion of the constant offset leads to pronounced window dependence, with the fitted $\gamma$ fluctuating strongly with $x_{\rm min}$ and the asymptotic behavior becoming unreliable at large distances. On this basis, we use the pure power-law form to characterize the long-range decay of the OBDM.

\section{One-Body Density Matrix for Soft-Core Bosons}
\label{secS3:Softcore}
\begin{figure}[h]
\includegraphics[width=\linewidth]{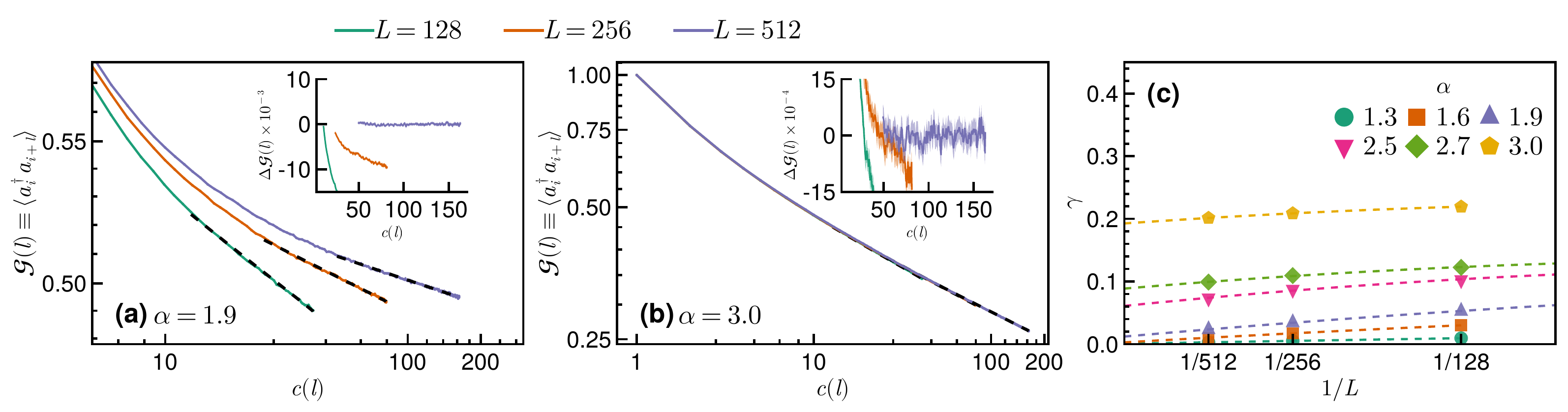}
    \caption{Characterization of the superfluid phase near MI-SF phase transition: (a)-(b) One-body density matrix, $\mathcal{G}(\ell)$, plotted against $\ell$ for $\alpha = 1.9$, and $3.0$. The dashed line represents the best fit to $A \cdot c(\ell)^{-\gamma}$, where $c(\ell) = \sin(\pi \ell / L)/ \sin(\pi/L)$ is the chord distance, and $A$ and $\gamma$ are fitting parameters. Insets: Difference $\Delta \mathcal{G}(\ell)$ between $\mathcal{G}(\ell)$ for system size $L$ and the numerical fit for $L = L_{\text{max}}$, plotted as a function of the chord distance $c(\ell)$. The difference indicates an upward trend of $\mathcal{G}(\ell)$ for increasing lattice size. (c) Finite-size scaling of the power-law exponent $\gamma$ to the thermodynamic limit.
    }
    \label{fig:SoftCoreOBDM}
\end{figure}

\begin{figure}[h]
\includegraphics[width=\linewidth]{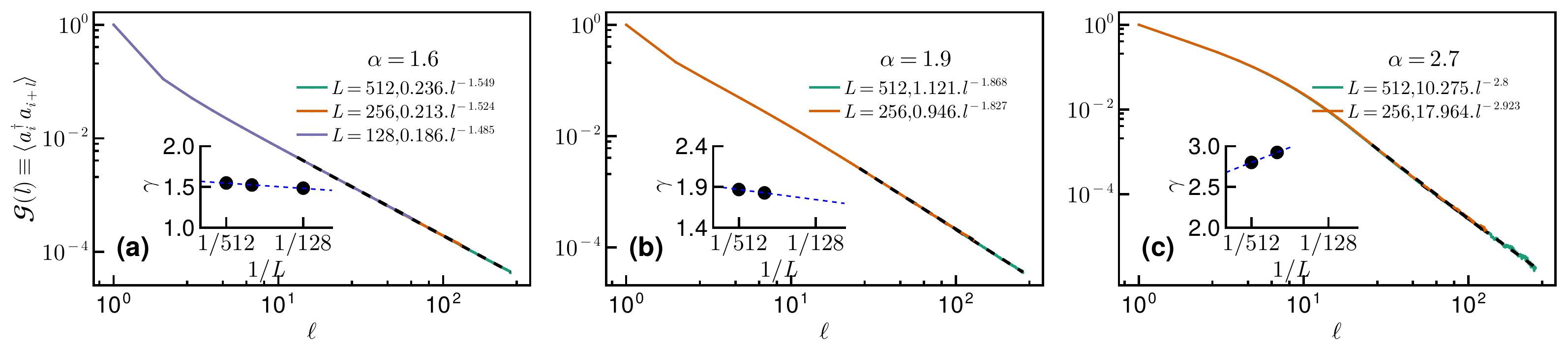}
    \caption{Characterization of the Mott-insulating (MI) phase via $\mathcal{G}(\ell)$ for $\alpha = 1.6, 1.9,$ and $2.7$, with lattice sizes up to $L = 512$. Panels (a)-(c) show $\mathcal{G}(\ell)$ decaying as a power-law, fitted to $A \cdot \ell^{-\gamma}$. Insets display the scaling of $\gamma$ with $1/L$, confirming convergence to $\alpha$ in the thermodynamic limit.
    }
    \label{fig:softcoreMI}
\end{figure}

In order to characterize the superfluid phase of a system of softcore bosons on a 1D lattice with power-law hopping, the one-body density matrix $\mathcal{G}(\ell)$ is evaluated for $1 < \alpha \leq 3$ on lattice sizes up to $L = 512$. Figure \ref{fig:SoftCoreOBDM} shows $\mathcal{G}(\ell)$ for selected values of $\alpha= 1.9,$ and $3.0$. Panels (a)-(b) in the figure show the data along with the best fits to the form $A \cdot c(\ell)^{-\gamma}$, where $c(\ell)$ is the chord distance, and $A$ and $\gamma$ are fitting parameters. For $\alpha = 1.9$, $\mathcal{G}(\ell)$ approaches a size-independent constant as L increases, indicating long-range order. In contrast, for $\alpha =3.0$, the correlations decay algebraically, consistent with quasi-long-range order. The insets in each panel display the difference $\Delta \mathcal{G}(\ell)$, defined as the deviation between $\mathcal{G}(\ell)$ for a lattice size $L$ and the numerical fit for the largest system size $L = L_{\text{max}}$. This difference, plotted against the chord distance $c(\ell)$, reveals an upward trend in $\mathcal{G}(\ell)$ as $L$ increases. Panel (c) illustrates the finite-size scaling of the power-law exponent $\gamma$ as the system approaches the thermodynamic limit. The fitted values of $\gamma$ are plotted against $1/L$, showing $\gamma \approx 0$ for $1 < \alpha \lesssim 2$, consistent with long-range order, and $\gamma > 0$ for $2 \lesssim \alpha \leq 3$, consistent with quasi-long-range order.

In the Mott-insulating phase, $\mathcal{G}(\ell)$ is analyzed for the same range of $\alpha$. Figure \ref{fig:softcoreMI} presents results for $\alpha = 1.6, 1.9,$ and $2.7$. The dashed lines in each panel represent fits to the form $A \cdot \ell^{-\gamma}$, where $A$ and $\gamma$ are the fitting parameters. In this regime, the correlation function exhibits an algebraic decay for all values of $\alpha$ studied. The extracted exponent $\gamma$ increases systematically with $\alpha$ and converges to the hopping exponent in the thermodynamic limit. The insets illustrate the finite-size flow of $\gamma$ as a function of $1/L$, confirming this convergence.

\section{BKT Benchmark: Winding Number Fluctuations in Short-Range Models}
\label{secS4:BKT}
\begin{figure}[h]
    \includegraphics[width=\linewidth]{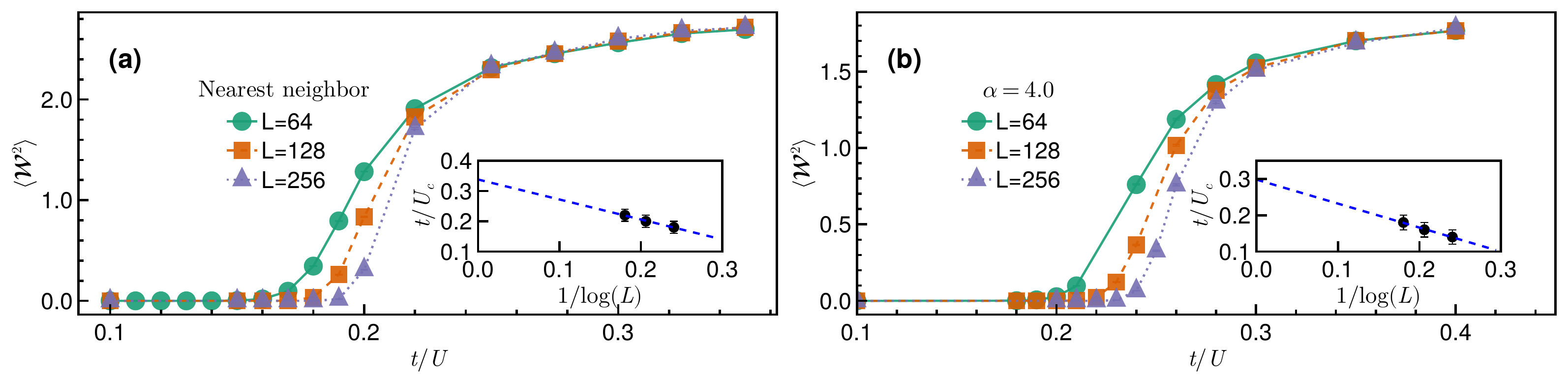}
    \caption{Mean-square winding number $\langle \mathcal{W}^2 \rangle$ as a function of hopping-to-interaction ratio $t/U$ probing the Mott insulator to superfluid quantum phase transition along the commensurate (unit-filling) density line for system sizes $L = 64, 128, 256$. (a) System with standard nearest-neighbor hopping, (b) Corresponding system with power-law hopping with decay exponent $\alpha = 4.0$. Insets show finite-size scaling of the critical point $t/U_c$ versus $1/\log(L)$, with dashed blue lines indicating extrapolation to the thermodynamic limit.}
     \label{fig:WindingSR_BKT}
\end{figure}
For effectively short-range hopping $(\alpha > 3)$, the superfluid–Mott insulator transition is expected to belong to the Berezinskii–Kosterlitz–Thouless (BKT) universality class. In this regime, the correlation length diverges with an essential singularity, which leads to Weber–Minnhagen finite-size scaling \cite{weberMonteCarloDetermination1988} of the critical coupling, $(t/U)_c(L) - (t/U)_c \propto 1/(\ln L)^2$. As a consequence, the winding-number fluctuations $\langle W^2 \rangle(t/U)$ do not exhibit a system-size-independent crossing, but instead display a characteristic drift of the apparent transition point with increasing system size. This behavior is shown in Fig. \ref{fig:WindingSR_BKT} for (a) the nearest-neighbor Bose-Hubbard model, and (b) power-law hopping model at $\alpha = 4$ (effectively short-range), where the rise of $\langle W^2 \rangle$ shifts with $L$, and the extrapolation of $(t/U)_c(L)$ versus $1/\log L$ is consistent with the expected BKT form. These short-range benchmarks provide a direct contrast with the scaling behavior observed in the long-range regime $1 < \alpha < 3$, where we instead observe a stable crossing and power-law finite-size scaling, indicating a universality class distinct from BKT.

\section{Experimental Realizations of Long-Range Hopping}
\label{secS5:Experiments}
Several experimental platforms implement Bose–Hubbard models with power-law hopping $J(r)\propto 1/r^\alpha$, or their hardcore/XY counterparts, through distinct microscopic mechanisms.

Excitonic lattices based on indirect dipolar excitons in GaAs bilayers realize intrinsic long-range hopping: each lattice site hosts a bosonic exciton mode whose electromagnetic field extends over multiple sites, leading to virtual-photon–mediated exchange and an experimentally observed hopping profile $J(r)\propto r^{-1.6}$ \cite{arxivLagoin2024BHLRH}. The effective range and strength can be controlled via lattice spacing, dipole moment, and photonic environment.

In trapped-ion chains, long-range flip–flop exchange is generated by phonon-mediated processes: Raman lasers couple internal states of ions at lattice sites $i$ and $j$ to delocalized vibrational modes, producing an effective XY coupling $J_{ij}\propto |i-j|^{-\alpha}$. The exponent $\alpha$ is tunable between nearly all-to-all and $\alpha\simeq 3$ by adjusting laser detunings and mode structure. Under the hardcore-boson mapping this corresponds directly to long-range hopping and has been experimentally demonstrated \cite{Richerme2014, Jurcevic2014, Monroe2021, LewisIonTrapLongrange2023}.

Rydberg atoms and polar molecules implement XY dynamics through resonant dipolar exchange, where pair states $|\alpha_i\beta_j\rangle$ and $|\beta_i\alpha_j\rangle$ are mixed at first order, yielding $J(r)\propto r^{-3}$ \cite{saffmanQuantumInformationRydberg2010, DefenuLongRange2023}. In Rydberg arrays this arises from Förster-enhanced exchange \cite{Barredo2015, browaeys_lahaye_2020, emperaugerTomonagaLuttingerLiquidBehavior2025, steinertSpatiallyTunableSpin2023, chen2023continuous}, while in polar molecules it originates from resonant exchange of rotational excitations \cite{Yan2013, hazzard2014, christakis2023}.

In superconducting circuits, long-range hopping is achieved using transmon qubits in photonic-bandgap metamaterial waveguides. A qubit tuned into the bandgap forms a qubit–photon bound state with exponentially decaying tails; overlap between neighboring sites produces coherent exchange $J_{ij}\propto e^{-|i-j|d/\xi}$ with controllable range \cite{sundaresan2019, zhangSuperconducting2023}.

\bibliography{references}